# Photoassisted tunneling from free-standing GaAs thin films into metallic surfaces

D. Vu,[a] S. Arscott,[b] E. Peytavit,[b] R. Ramdani,[c,d] E. Gil,[c,d] Y. André,[c,d] S. Bansropun,[e] B. Gérard,[f] A. C. H. Rowe,[a] D. Paget.[a+]

[a] Laboratoire de Physique de la Matière Condensée, Ecole Polytechnique CNRS, 91128 Palaiseau Cedex, France.

[b] Institut d'Electronique, de Microélectronique et de Nanotechnologie (IEMN), University of Lille, CNRS, Avenue Poincaré, Cité Scientifique, 59652 Villeneuve d'Ascq, France.

[c] Clermont-Université, Université Blaise Pascal, LASMEA, BP 10448, F-63000 Clermont-Ferrand, France.

[d] CNRS, UMR 6602, LASMEA, F-63177 Aubière, France.

[e] THALES Research and Technology France, Route départementale 128, 91767 Palaiseau cedex, France.

[f] ALCATEL-THALES III-V lab, THALES Research and Technology France, Route départementale 128, 91767 Palaiseau Cedex, France.

Abstract.

The tunnel photocurrent between a gold surface and a free-standing semiconducting thin film excited from the rear by above bandgap light has been measured as a function of applied bias, tunnel distance and excitation light power. The results are compared with the predictions of a model which includes the bias dependence of the tunnel barrier height and the bias-induced decrease of surface recombination velocity. It is found that i) the tunnel photocurrent from the conduction band dominates that from surface states. ii) At large tunnel distance the exponential bias dependence of the current is explained by that of the tunnel barrier height, while at small distance the change of surface recombination velocity is dominant.

I Introduction

Spin injection from GaAs under light excitation into a magnetic metal is of fundamental interest for spintronics[1] and spin-polarized scanning tunnelling microscopy.[2] In contrast with injection from magnetic tips,[3] the use of semiconducting injectors permits rapid (optical) control of the spin of the injected electrons and minimizes the magnetic interactions between the injector and the surface. Some attempts at spin injection from GaAs tips into magnetic surfaces have already been made[4,5] but these studies used direct light excitation of the tip apex and a parasitic dependency of the injected current on the light helicity as high as several percent was observed.[6] This deleterious effect, attributed to helicity-dependent light scattering in the tunnelling gap, seriously limited the use of the GaAs-tip injectors. It has since been proposed that spin injectors should operate in transmission mode, with light excitation incident on the planar back surface of the injector.[7, 8]

In order to understand the features of spin injection it is first necessary to understand the mechanisms of charge injection via tunnelling from a photo-excited semiconductor into a metal. To our knowledge, despite the large number of studies of photoelectric effects in metal-semiconductor junctions[9, 10] and tunnel microscopes,[11] a complete understanding of photoelectrical processes is still lacking. A previous study using silicon tips found that the dominant process is a Fowler-Nordheim-like one.[12] Jansen *et al.* proposed that electrons tunnel from midgap surface states thereby obtaining good agreement with experimental results for small values of the bias applied to the metal.[13, 14] These studies considered an energy-independent density of surface states and neglected surface recombination as well as the bias-dependent tunnel barrier height. Injection of free carriers across a semiconductor-liquid interface has also considered,[15] with tunnelling from the conduction or valence band accounted for. However, in this case the only applied bias was the constant photovoltage.

In this work the tunnel photocurrent into a gold surface is measured as a function of bias, film/surface distance and light excitation power. The GaAs film is a free-standing cantilever having a thickness of a few micrometers.[16] We consider thin GaAs films photoexcited from the rear face and held at a controlled distance from the metal. This configuration brings two simplifications to the understanding of the results: i) Since light excitation is performed from the rear of the film, the injected photocurrent originates from electrons created near this surface which have diffused to the front surface. Unlike front surface excitation,[17] this photocurrent does not directly depend on the width of the depletion layer. ii) The use of a film rather than a tip or a sharp point ensures that the contact surface is relatively large thereby avoiding the effects of a complex electric field distribution near the tip apex.[13] Analysis of the results has allowed us to eliminate the effect of possible distance inhomogeneities so that, after correction, the metal-semiconductor interface can be considered as planar, in the sense of a parallel plate capacitor.

The results are analyzed using a new model which incorporates photovoltage,[10] surface recombination[18] and the energy dependence of the density of surface states[19] together with the bias dependence of the tunnel barrier height.[20] For a gold (non-magnetic) surface, the interpretation of the results is relatively simple as the density of empty states depends only weakly on energy.[21] The quantitative agreement between the model and the experimental data demonstrates that, unlike earlier work,[13] the tunnel photocurrent originates from the conduction band. For a large tunnel distance, the observed exponential relationship of the tunnel photocurrent is caused by the bias dependent tunnel barrier height. At short distances, the dependence becomes non-exponential and is determined by the bias-induced change of the surface recombination velocity.

This paper is organized as follows. Sec. II details the model while the experimental results and procedure are presented in Sec. III. Comparison between the model and the experimental data is found in Sec. IV whilst Sec. V presents a general discussion.

II Theory

The metal-semiconductor structure, described in Fig. 1, is composed of a p-type semiconductor film (thickness $\ell$, bandgap $E_G$) and a metal to which a potential $V$ is applied, separated by an insulating layer of thickness $d$ and dielectric constant $\varepsilon_t$. Light excitation from the rear of the semiconductor creates a population of photoelectrons in the conduction band, with a fraction of these electrons being injected into the space charge region at the interface.

A Surface photovoltage and surface recombination velocity

Neglecting the difference between the Fermi energy in the semiconductor and the top of the valence band, the potential barrier at the semiconductor surface, defined by the energy difference between the top of the valence band in the bulk and at the surface, is given by

$$\varphi_b = \varphi_0 + \Delta\varphi - qV_s. \tag{1}$$

Here $qV_s$ is defined as the energy difference between the electron quasi Fermi level at the surface and the bulk Fermi level, caused by light excitation and by the application of a bias. ($q$ is the negative electronic charge). The energy $\Delta\varphi$ is the shift of the electron quasi Fermi level with respect to midgap caused by the change of the surface charge. While a general calculation can readily be performed, it will be assumed that the equilibrium value of the surface barrier $\varphi_0$ is equal to half the bandgap energy and that the density of surface states peaks at midgap.

In order to calculate $\Delta\varphi$, $qV_s$ and the electron concentration $n_0$ at the onset of the depletion region three conservation equations are used. The first is the charge neutrality equation,

$$\delta Q_m + \delta Q_{sc} + \delta Q_{ss} = 0, \qquad (2)$$

where the three terms are the departures from equilibrium of the charge densities at the metal, at the semiconductor surface and in the semiconductor depletion layer. The two conservation equations for the electron and hole current densities are:

$$J_p - J_{te} = J_r \qquad (3)$$

and

$$J_s = J_0 \exp\left(-\frac{\Delta\varphi}{kT}\right)\left[\exp\left(\frac{qV_s}{kT}\right) - 1\right] = J_r - J_{th}, \qquad (4)$$

where $J_p$ is the photocurrent density injected into the depletion region and $J_r$ is the current density for electron-hole surface recombination. The tunnel current densities $J_{te}$ and $J_{th}$ describe electron tunnel processes from the semiconductor to the metal and hole processes from the metal to empty states of the semiconductor, respectively. $J_s$ is the majority carrier Schottky current.[9] Here $J_0 = A^{**}T^2 \exp\left(-\frac{\varphi_0}{kT}\right)$ is the usual saturation current density where $A^{**}$ is the effective Richardson constant, $T$ is the temperature and $k$ is the Boltzmann constant.

The calculation of the tunnel current can be simplified if $J_{te} \ll J_p$. This assumption is valid provided the tunnel gap is not too small and will be justified below by comparison with the experimental results. In the opposite extreme case, the tunnel photocurrent is equal to the injected photocurrent ($J_p$) and the photovoltage is small. Similarly, it will be assumed that $J_{th} \ll J_s$.

As shown in Appendix A, the expressions for $n_0$ and for $J_p$, obtained from a resolution of the one-dimensional diffusion equation in the semiconductor bulk, are

$$n_0 = \beta N_0 \tag{5}$$

and

$$J_p = qN_0 S\beta \tag{6}$$

respectively, with

$$\beta = (1 + S/v_d)^{-1}. \tag{7}$$

The effective electron concentration $N_0$ is proportional to the light excitation power and the diffusion velocity $v_d$ is proportional to the ratio of diffusion constant and diffusion length. Neither quantity depends on the surface recombination velocity $S$ or bias. Their expressions are given in Appendix A.

Using $J_p = J_s$ and further assuming in Eq. (4) that $e^{\frac{qV_s}{kT}} \gg 1$, Eq. (6) becomes

$$qV_s = qV_s^* + \Delta\varphi + kT\ln(1-\beta). \tag{8}$$

The quantity $V_s^*$, defined by $qV_s^* = kT\ln(qv_d N_0 / J_0)$ is the usual value of the photovoltage ($kT\ln(J_p/J_0)$) in the limit where $S \gg v_d$. With respect to most studies performed using light excitation at the front surface,[10] the transmission geometry strongly simplifies the expression for the surface recombination dependence of the photovoltage. Assuming thermodynamic equilibrium between bulk and surface, it is straightforward to calculate the electron concentration $n_s$ at the surface. This concentration mostly lies at the energy of the lowest quantized state in the surface depletion layer. Here this energy lies above the bottom of the conduction band and is written $f^*\varphi_b$ where[22]

$$f^* \approx \frac{1}{\varphi_b}\left(\frac{q^2\hbar^2 E_{eff}^2}{2m^*}\right)^{1/3}[3\pi/4]^{2/3} \tag{9}$$

and $E_{eff}$ is the surface electric field. Assuming that the quasi Fermi level position is independent of space[10, 23] and using Eq. (4) and Eq. (5) one finds

$$n_s = \frac{A^{**}T^2}{qS}\left[\frac{qSn_0}{A^{**}T^2}\right]^{f^*}. \tag{10}$$

The surface recombination velocity is obtained from Eq. (3). Neglecting recombination in the depletion layer and considering the usual Stevenson-Keyes expression for $J_r$,[18, 24] this equation is written

$$N_0 S\beta = \int_{E_{Fh}}^{E_{Fn}} N_T(E) \frac{\sigma_n \sigma_p v_n v_p n_i^2 \left(e^{qV_s/kT}-1\right)}{\sigma_n v_n (n_s+n_{ts}) + \sigma_p v_p (p_s+p_{ts})} dE \tag{11}$$

It is assumed that the density of surface states $N_T(E)$ has a maximum $N_T(0)$ at midgap and a typical width 0.2 eV.[19] Here, $\sigma_n$ and $\sigma_p$ are the electron and hole capture cross sections, of respective velocities $v_n$ and $v_p$ and $n_i$ is the intrinsic electron concentration. The quantities $p_s$, $n_{ts}$ and $p_{ts}$ are respectively the surface hole concentrations and the values that $n_s$ and $p_s$ would have if the surface Fermi level were at energy $E$.

As bulk and surface are in thermodynamic equilibrium, the second term in the denominator of Eq. (11) is generally smaller than the first one. This also implies that hole recombination processes are less efficient than electron ones and that the occupation probability is close to unity for all states lying between the two quasi Fermi levels.[25] As a result, the only states which contribute to surface recombination are in a relatively narrow range of typical width $kT$ situated near $E_{Fn}$. Using the standard room temperature value of the intrinsic density of states of the conduction band, one finds that $n_s \gg n_{ts}$, so that

$$S = S_0 \exp(-\frac{\Delta\varphi}{kT})/D(\Delta\varphi), \tag{12}$$

where $D(\Delta\varphi)$ is the relative surface state density at energy $\Delta\varphi$. The equilibrium surface recombination velocity is given by $S_0 = (J_0^2 / qN_T^* J_{r0}) e^{\frac{\varphi_0}{kT}}$, where $N_T^* = N_T(0)kT/a$ is an equivalent volume concentration of defects and $J_{r0} = qv_p n_i (an_i \sigma_p)$. The thickness $a$ of the surface only plays a role for the homogeneity of $N_T^*$ and $J_{r0}$.

Expressing $\delta Q_m$ using Gauss's theorem, $\delta Q_{ss}$ by an integration on surface states, and taking account of the contribution to $\delta Q_{sc}$ of conduction electrons,[10] the charge neutrality equation, Eq. (1), becomes

$$C_m \left[ V - \frac{\varphi_b - \varphi_0}{q} \right] + qW_0 N_A \left[ \sqrt{\left[\frac{\varphi_b}{\varphi_0}\right]\left[1 + \frac{n_s}{N_A}\frac{kT}{\varphi_b}\right]} - 1 \right] + qN_T(0) \int_{E_{Fh}}^{E_{Fe}} D(\varepsilon) d\varepsilon = 0, \quad (13)$$

where $W_0$ is the equilibrium value of the depletion layer width, $N_A$ is the acceptor concentration, and $C_m = \varepsilon_t / d$ is the capacitance per unit area of the tunnel gap. Since $qV_s$ and the $S$ are expressed as a function of $\Delta\varphi$, [Eq. (8) and Eq. (12) respectively] this energy is the relevant quantity for calculating the tunnel currents and is found by numerical solving Eq. (13).

B Calculation of the photoassisted tunnel currents

The tunnel current density $J_t$ is generally the sum of three contributions describing respectively tunnelling of photoelectrons from the conduction band ($J_{tb}$), from surface states ($J_{ts}$), and of the current from the valence band ($J_{tv}$). Electron tunnelling between the conduction band and the metal occurs by conservation of the parallel electronic momentum and of the total energy.[26] In addition to the electronic perpendicular momentum, the tunnel probability depends on the spatial average of the tunnel barrier, which itself depends on bias.[20] As discussed in Appendix B, the tunnel probability is a function of energy above the conduction band edge. The majority of tunnelling electrons have a nonzero energy, written $f\varphi_b$ where $f \geq f^*$ is a number *a priori* distinct from $f^*$ and defined by the lowest quantized state [Eq. (9)]. For $J_{tb}$ one then obtains

$$J_{tb} = J_{tb}^0 N^*(S) \exp\left(-\frac{\omega qV}{kT}\right) \tag{14}$$

where

$$\omega = \frac{d}{2d_0} \frac{kT}{\sqrt{\Phi_b^*}}, \tag{15}$$

$$N^*(S) = \frac{A^{**}T^2}{qS} W_0\left[\frac{qSn_0}{A^{**}T^2}\right]^{f+\omega(1-2\alpha f)} \tag{16}$$

and

$$J_{tb}^0 = K_b \rho_m [E] \exp\left[-\frac{\omega(\Phi_b^*/4 - (1-2f)\varphi_0)}{kT}\right] \tag{17}$$

are respectively the reduced distance, the surface electron concentration and a factor independent of bias and light excitation. The other quantities are $d_0 = \hbar/\sqrt{2m}$, $\Phi_b^* = [\Phi_m + \chi - E_G + (1-2f)\varphi_0]/2$ and $E = E_g - (1-f)\varphi_b + qV$, the energy of tunnelling electrons with respect to the metal Fermi level. The quantity α, defined in Eq. (B4), is the fraction of perpendicular to total kinetic energy. $K_b$, defined by Eq. (B5) is a constant. The exponential factor in Eq. (14) is due to the bias dependent tunnel barrier, while the bias dependence of $N^*(S)$ reflects the changes of the surface recombination velocity.

The tunnel current from surface states is obtained by integration over energy ε with respect to midgap between the electron quasi Fermi level and the metal Fermi level. One finds

$$J_{ts} = N_T(0) A \exp\left\{-\frac{\omega_s}{kT}[q(V-V_s) + \Delta\varphi]\right\} \int_{\Delta\varphi+q(V-V_s)}^{\Delta\varphi} K(E_s) \rho_m(E_s) D(\varepsilon) \exp(\frac{2\varepsilon\omega_s}{kT}) d\varepsilon \tag{18}$$

where $\Phi_s^* = \frac{1}{2}(\Phi_m + \chi + \varphi_0)$,    $\omega_s = \frac{d}{2d_0}\frac{kT}{\sqrt{\Phi_s^*}}$,    $A = \exp\left[-2d\sqrt{\Phi_s^*}/d_0\right]$    and

$E_s = \varepsilon - \Delta\varphi + qV_s$. Taking account of Eq. (8), this expression becomes

$$J_{ts} = N_T(0)A\exp\left[-\frac{q\omega_s V}{kT}\right]\left[\frac{qSn_0}{J_0}\right]^{\omega_s}\int_{\Delta\varphi+q(V-V_s)}^{\Delta\varphi}K_s(E)\rho_m(E)D(\varepsilon)\exp(\frac{2\varepsilon\omega_s}{kT})d\varepsilon. \quad (19)$$

In this expression it is noted that the dependence on light excitation power is mostly contained in the term $[qSn_0/J_0]^{\omega_s}$.

The tunnel current from the valence band can also be modulated under light excitation since the photovoltage modulates the energy of the top of the valence band at the surface. This equivalent photocurrent appears as soon as $qV > \varphi_b$ and is given by

$$J_{tv} = J_{tv}^0 \exp\left[-\frac{\omega_v qV}{kT}\right]\left[\frac{qSn_0}{J_0}\right]^{\omega_v}$$
$$\int_0^{qV-\varphi_b} G_v(\varepsilon_v)\rho_m(E)D(\varepsilon_v)\exp(-\frac{4\omega_v \varepsilon_{v\perp}}{kT})d\varepsilon_v \quad (20)$$

where $\omega_v = \frac{d}{2d_0}\frac{kT}{\sqrt{\Phi_v^*}}$, $\Phi_b^* = [\Phi_m + \chi + E_G + \varphi_0]/2$ and $J_{tv}^0 = K_v \exp\left[-\frac{2d\sqrt{\Phi_v^*}}{d_0}\right]$. The density of states per unit surface is $D(\varepsilon_v) = \ell_c(2m^*/\hbar^2)^{3/2}\sqrt{\varepsilon_v}$, where $\ell_c$ is the coherence length. $K_v$ gives a measure of the tunnel matrix element and $G(\varepsilon_v)$ is a slowly varying function similar to that defined in Appendix B for conduction electrons. In the same way as for surface states, the power dependence of this current is given by the third factor of Eq. (19) and is of the form $N_0^{\omega_v}$ where $\omega_v$ is smaller than $\omega_s$ because of the larger value of the tunnel barrier.

.

III Experimental

A Experimental system and procedure

We have used free-standing, 3 μm thick, cantilever patches of $p^+$ GaAs (doping $\approx 10^{18} cm^{-3}$) deposited on fused silica substrates with the cantilever overhanging the substrate. These devices have been fabricated using an original microfluidic assembly process developed by some of the authors.[16] No preliminary surface treatment was given to the cantilevers before the experiment. The cantilevers were excited by a laser diode at 1.59 eV, of power 5 mW, focussed to a spot of 20 μm diameter. The laser beam reflected by the cantilever was also detected by a quadrant photodiode which permits the measurement of the force between cantilever and the surface and therefore to characterize the mechanical contact. Freshly made, atomically smooth, Au surfaces, fabricated using an electrochemical technique described elsewhere,[27] were used for the experiments.

As described in Ref. (12), the current was stabilized in the dark to a value $I = 10 nA \, x \exp(I_{set}/2000)$ for a cantilever bias $V_{set}$ (set here to -1.5V). The value $I_{set}$ is adjusted using the feedback control system between values situated between -3000 and +3000 and determines the tunnel distance if quantities such as the dielectric constant of the tunnel gap are constant. After stabilization, the feedback loop was opened and two rapid bias scans were performed, one in the dark and the other one under illumination. This procedure allows us to measure both the dark current $I_{dark}$ and the additional tunnel current $I_{ph}$ due to light excitation as a function of bias.

In the following we show the bias dependences as a function of $I_{set}$ rather than of the tunnel distance which is not known accurately. Fig. 2 shows the dark current as a function of bias, while Fig. 3 shows the absolute value of the additional photocurrent. For a positive bias, the photocurrent has the opposite sign and is due to tunnelling of holes to occupied states of the metal. This process, which will not be discussed here, compensates the electron photocurrent at a bias of about 0.2 eV which is therefore not directly related to the standard photovoltage. Fig. 4 shows the atomic force between the cantilever and the metal surface. Finally, the photocurrent as a function of light excitation power, for an applied bias of -1.5 V is shown in Fig. 5. This figure shows a power law with an exponent which slightly increases with distance from 0.44 (Curve d) to 0.66 (Curve a).

B Analysis

The experimental results of Fig. 1, Fig. 2 and Fig. 3 allow us to distinguish two regimes as a function of $I_{set}$. As seen in Fig. 2, for $I_{set} \leq 0$, the photocurrent behavior is very close to exponential, with a slope which decreases with distance between Curve a and b and increases again between Curves b and c. For larger values of $I_{set}$ the photocurrent increases more slowly than exponential. The limit between the two behaviors approximately coincides with the onset of mechanical contact which, as seen in Fig. 4, occurs between $I_{set} = 0$ and $I_{set} = 1000$. In forward (positive) bias, it is possible to define an ideality factor $n$ since the dark current exhibits exponential behavior according to $\exp(qV/nkT)$. Fig. 2 shows that for $I_{set} < 0$, in agreement with Ref. (32), the ideality factor decreases with increasing $I_{set}$. For $I_{set} \geq 0$ the slope of the exponential is constant which shows that the capacitance $C_m$ of the tunnel gap is constant. The overall variation of the ideality factor is from 2.7 to 1.5.

These results can be given a simple explanation, summarized in the inset of Fig. 4: before mechanical contact the tunnel distance and therefore the ideality factor decrease with increasing $I_{set}$. Once mechanical contact is established the bias dependence of the tunnel photocurrent becomes a sum of a contact contribution, characterised by a fixed distance, together with a non-contact contribution. The relative importance of each contribution depends on the ratio of the two areas and thus on $I_{set}$. Support for this hypothesis is shown in Fig. 6 where for $I_{set} \geq 0$, the photocurrent and the dark current are decomposed as the sum of a contribution independent on $I_{set}$ and of a fraction α of the signal obtained at $I_{set} = -1000$. This is the highest $I_{set}$ value giving an exponential bias dependence of the tunnel photocurrent. In reverse (negative) bias, both the tunnel photocurrent and the dark current bias dependences are nearly independent of $I_{set}$. Further, the dark current is now exponential as a function of forward (positive) bias over as much as 4 orders of magnitude. The value of $\alpha$ is given in Table 1. This value gives a measure of the relative area of the non-contact part to the contact one and, as expected, decreases upon increasing $I_{set}$. In contact it is interesting to note that a bistability of the atomic force is observed (see Fig. 4). This bistability is correlated with a bistability of the tunnel photocurrent and can be corrected in the same way as above using two distinct values of $\alpha$ as shown in Table 1 for $I_{set} = 3000$. For the following analysis only the smallest value of $\alpha$ will be considered for each value of $I_{set}$.

The corrected results are summarized in Fig. 7 which shows the bias dependence of the tunnel photocurrent in the contact regime and in the non contact regime as a function of $I_{set}$. The ideality factor, also shown in Fig. 7, increases from a nearly constant value of 1.5 in contact to 2.7 at large distances. These results are free of possible contact inhomogeneities arising from the large contact surface area and reveal the tunnel characteristics of a purely two-dimensional contact considered in Sec. II.

IV Interpretation

Since the model described in Sec. II contains a relatively large number of parameters, we have chosen reasonable values of several parameters from the literature. These are given in Table 2 and no attempts have been made to adjust them. The values of the surface recombination velocity $S_0$, the diffusion constant $D$, the diffusion length $L$ and the bulk recombination time, $\tau$, are summarized in Ref. (8). The values of $N_0$ and $v_d$, calculated using Eq. (A3) and Eq. (A4) were taken equal to 2 x$10^{22}$ m$^{-3}$ and 1600 m.s$^{-1}$ respectively. The energy dependence $D(\Delta\varphi)$ of the density of surface states is approximated by a Gaussian profile of width $\sigma$, estimated to be 0.20 eV,[19, 28] whereas for $\Delta\varphi$ larger than $\sigma$, the tails of the valence and conduction bands are approximated by parabolas. The density of surface states $N_T(0)$ has been found to range from several $10^{17}$ eV$^{-1}$m$^{-2}$ up to larger than $10^{18}$ eV$^{-1}$m$^{-2}$.[19, 28] Here we take $N_T(0) = 6\times10^{18}$ eV$^{-1}$m$^{-2}$ as implied by the slopes of the bias dependences at large distance and discussed in Sec. IVB below. Since $C_m$ and $\omega$ depend on the width of the tunnel gap, their values are adjusted for each spectrum whilst maintaining constant values of $\Phi_b^*$, and $\Phi_s^*$. Using Eq. (16), we take the exponent of the experimental power dependence of the photocurrent for the factor $f$ ($\approx 0.4$). For simplicity we take $f^* = f$, thus assuming that tunnelling of photoelectrons occurs from the first quantized state in the depletion layer.

In the following, we outline the physical mechanisms underlying the tunnel photocurrent from a semiconductor into a metal.

A Tunnel currents from surface states and from the conduction band.

The relative values of $J_{ts}$, $J_{tv}$ and $J_{tb}$ depend on the tunnel matrix elements which are unknown, so that the magnitudes of these currents cannot be conclusively determined. However, the experimental evidence presented here is at variance with the model of Jansen et al.[13] in that the tunnel photocurrent from surface states and from the valence band are negligible with respect to that from the conduction band.[29] This is most apparent for the following two reasons:

a) The predicted excitation power dependences of $J_{ts}$ and $J_{tv}$ are very weak and cannot explain the experimental results. Recalling that $\beta \ll 1$ at large distance, it is seen from Eq. (19) and Eq. (20) that these dependences are dominated by that of $N_0^{\omega_s}$ and $N_0^{\omega_b}$. The exponents $\omega_s$ and $\omega_b$ are of the order of $1 \times 10^{-2}$ and are more than one order of magnitude smaller than the experimental values. Even larger discrepancies are found in contact.

b) The bias dependences of $J_{ts}$ and $J_{tv}$ cannot interpret the data. This is shown in Fig. 8 for the extreme case of Curve a and Curve d in Fig. 7. The bias dependence of $J_{tv}$ exhibits a threshold near -0.4 V and nonexponential behavior which does not interpret the data at large distance. Conversely, because of the nonlinear integral of Eq. (18), $J_{ts}$ is almost independent of bias in contact and cannot interpret the experimental data. Even a strong modification of the tunnel parameters cannot account for the experimental results.

B Tunnel current from the conduction band

Comparison of the experimental results with the bias dependences of $J_{tb}$, calculated using Eq. (14) are shown in Fig. 7. Very good agreement with the experimental results is obtained. The values of $C_m$ and $\omega$ used in the comparison are given in Table 2. Both of them increase with decreasing $I_{set}$, which reveals an increase of the tunnel distance. Fig. 9 shows Curves $a$ and $d$ from Fig. 7 along with the calculated bias dependences of $N^*(S)$ and $\exp(-\omega qV/kT)$ which appear in Eq. (14). While the exponential factor accounts for the bias dependence of the tunnel barrier height, $N^*(S)$ expresses the bias-induced decrease of the surface recombination velocity which, according to Eq. (9), produces an increase of the concentration of tunnelling electrons.

At large distance, the bias dependence of the tunnel current is due to that of the tunnel barrier height as the surface recombination velocity only weakly depends on bias. Indeed, Eq. (12) simplifies into

$$\Delta \varphi \approx -\gamma_t^* q[V - V_s^*] = \frac{C_m}{q^2 N_T(0)} q[V - V_s^*]. \tag{21}$$

$\Delta\varphi$ is smaller than the width $\sigma$ of the surface density of states so that the electron quasi Fermi level is still pinned near midgap. The linear bias dependence of $\Delta\varphi$ induces an exponential dependence of the tunnel photocurrent, proportional to $\exp(-V/V_{ph})$, where

$$\frac{kT}{qV_{ph}} = \gamma_t^* + \omega. \qquad (22)$$

The second term of Eq. (22), given by Eq. (16), is proportional to $d$ and expresses the bias dependence of the tunnel barrier. The first term, which is proportional to $d^{-1}$, reflects the dependence of the tunnel barrier on $\Delta\varphi$. The observed decrease of the slope for $I_{set}$ increasing between -2500 and -2000 implies that the exponential increase of the tunnel current is determined by the bias dependence of the tunnel barrier and that $\gamma_t^* < \omega$. The subsequent increase between -2000 and -1000 suggests that $d$ is now small enough so that $\gamma_t^* > \omega$. The condition $\gamma_t^* \approx \omega$ implies that the value of $\omega$ is given by the measured exponential slope at large distance. Using the values of $C_m$ and $\omega$ given in Table 2, one finds that $qN_T(0)$ should be of the order of several $10^{18}$ eV$^{-1}$.m$^{-2}$ which is indeed the case.

At small distances, $\Delta\varphi$ increases because of the larger value of the tunnel capacitance $C_m$. $\Delta\varphi$ can become larger than the width $\sigma$ of the band of surface states which induces an unpinning of the surface Fermi level and a decrease of the surface recombination velocity. The bias dependence of the tunnel current is now caused by the increase of the electron concentration $n_s$ which, as seen in the top panel of Fig. 9 is as large as three orders of magnitude.

V Discussion

       A Effects of interface chemistry

The values of the parameters used in the analysis suggest that the natural oxide layer, originally present at the surface, has been at least partially removed. For a Schottky barrier composed of gold deposited on naturally-oxidized GaAs, one finds a value of $\varepsilon_0/C_m = d/\varepsilon_t \approx 1.5$, about 2 orders of magnitude larger than the one measured here in contact.[30] As shown for InP, the oxide may have been removed by an electrochemical reaction at cathodic potentials.[31]

Taking $\Phi_b^* \approx 4$ eV in Eq. (15) one finds that the distance $d$ ranges between 1.1 nm to 0.45 nm in the non contact regime and is about 0.28 nm in contact. The resulting values of the dielectric constant of the interfacial layer $\varepsilon_t$ are shown in Fig. 10 as a function of distance. $\varepsilon_t$ is equal to $\varepsilon^* \approx 30$ in contact which suggests the partial formation of a molecular film of water (dielectric constant 80 and thickness $d^* \approx 0.28$ nm) between the semiconductor and the metal. If $d > d^*$, one expects the effective dielectric constant to be given by $\varepsilon_t = d\varepsilon^* [d^* + (d - d^*)\varepsilon^*]^{-1}$. As shown in Fig. 10 for the non contact regime, the correspondence between the calculated dependence of $\varepsilon_t$ and the data is unexpectedly good, given the uncertainties in most parameters used in the calculation.

B Dark current

For a forward (positive) biases, including the contribution $\alpha_d$ of residual processes such as image charge effects and tunnelling of majority carriers, the ideality factor is given by [9, 32]

$$\frac{1}{n} = 1 - \frac{\varepsilon_s/W + q(1-\eta)N_T(0)}{C_m + \varepsilon_s/W + qN_T(0)} - \alpha_d \approx \eta - \alpha_d \qquad (23)$$

where $\eta$ and $1-\eta$ are the fractions of the total number of states for which follow the metal statistics and the semiconductor statistics respectively. Comparison of the model with the data as discussed in Sec. IV.B shows that $\varepsilon_s/W \ll q(1-\eta)N_T(0)$ and $C_m \ll qN_T(0)$ which leads to the approximate expression in Eq. (23). Fig. 10 shows the dependence of $n^{-1}$ on $\varepsilon_0/C_m$, where the value near zero corresponds to the contact situation.

Under reverse (negative) bias, $\Delta\varphi$ and $qV_s$ are found by numerically solving the current and charge conservation equations and the dark tunnel current from surface states is given by Eq. (18).[33] Current conservation implies that the tunnel and Schottky currents are equal. For the Schottky current, in order to take account of additional processes contributing to the ideality factor, one replaces $\varphi_0$ by $\varphi_0^*$ which depends on the barrier change $\Delta\varphi - qV_s$. For a large bias range we write to second order

$$\varphi_0^* = \varphi_0 + \alpha_d (qV_s - \Delta\varphi) + \alpha_d' (qV_s - \Delta\varphi)^2 \qquad (24)$$

In the charge neutrality equation, in addition to $n_s = 0$, the term $\delta Q_{ss}$ must take account of the two types of surface states used in forward (positive) bias. The dark current and ideality factor depend on the following additional parameters: i) $\eta$. ii) $\alpha_d$ and $\alpha_d'$ iii) The tunnel matrix element $K_s$ defined in Eq. (18). Since this equation uses the product $K_s N_T(0)$, the quantity $N_T(0)$ will be replaced by an effective density of states $N_T^d(0)$ taken here as $8 \times 10^{17}$ eV$^{-1}$.m$^{-2}$.

The dependences of the dark current under reverse bias were calculated using the same parameter values as for the photocurrent as well as imposing $\eta - \alpha_d = n^{-1}$ from Eq. (23). The dependences of $\eta$ and $\alpha_d$ on $\varepsilon_0/C_m$ are shown in Fig. 10. As shown in Table 2, $\alpha_d'$ is only significant in contact and has very small values of the order of $10^{-3}$ V$^{-1}$. The bias dependences of the dark current under reverse bias, shown in Fig. 7, accounts very well for the experimental results. The calculations also suggest that, as expected,[34] the quantity $\eta$ decreases with increasing distance from a value of about 0.84, while the residual ideality factor $(1-\alpha_d)^{-1}$ decreases from 1.20 to 1.04.

B Validity of the approximations made

It has been assumed that the electrons in the conduction band tunnel from the first quantised level ($f \approx f^*$). The power dependence of the tunnel photocurrent gives $f \approx 0.4$. The bias dependence of $f^*$ was calculated using Eq. (9), neglecting the modification of the surface electric field due to the photoelectrons in the depletion layer: $f^*$ is approximately constant and varies from 0.38 (a value close to $f$) to about 0.25 as a function of bias. In view of the numerous quantities which play a role in defining the value of $f$ it is concluded that taking $f \approx f^*$ is a valid approximation.

At small distances, image charge effects might further modify the bias dependence of the tunnel photocurrent.[20] However, the characteristic energy for evaluating the magnitude of these effects $\lambda = q \ln(2)/(8\pi\varepsilon_t d)$, of the order of 0.4 eV for $\varepsilon_t = \varepsilon_0$ and $d = 1$nm, is smaller by one order of magnitude than the effective tunnel barrier height $\Phi_b^*$. As seen in Ref. (20), the the tunnel barrier decreases with bias so that image charge effects should induce a super-exponential increase in the tunnel photocurrent. This is at odds with the experimental results obtained at small distance.

In order to obtain analytical expressions of the tunnel current, this current has been neglected with respect to the photocurrent and Schottky currents. This assumption is certainly valid at large distance, in which case the tunnel photocurrent is small. In contact, the photocurrent $J_p \approx qN_0 S/v_d$ decreases because of the reduced surface recombination velocity and could become a lower limit value for the tunnel photocurrent ($J_t = J_p$). However, the latter hypothesis can also be excluded because, in contradiction with the results of Fig. (5), the resulting power dependence of the tunnel current should be quite different from that obtained at large distance.

VI Conclusion

We have developed a general model for describing the bias and distance dependence of the tunnel photocurrent from a thin free-standing GaAs film photo-excited from the rear surface and a metallic surface. Based on current and charge conservation equations, this model predicts that the tunnelling current can depend on bias via the bias dependence of the tunnelling barrier and also because application of bias changes the position of the electronic quasi Fermi level at the surface and therefore the effective density of states for surface recombination. Both the tunnel currents from the conduction band and from surface states have been calculated.

This model was compared with experimental data of tunnelling injection into gold surfaces, for which the density of empty states depends only weakly on energy. All results, including tunnel photocurrent, ideality factor and dark current under reverse bias, are satisfactorily interpreted by identical values of the parameters, close to the values found in the literature. The values obtained for the width and dielectric constant of the tunnel gap are also reasonable. The model and experimental results indicate that:

- The dominant part of the tunnel photocurrent comes from the conduction band.
- At large distance, the bias dependence of the tunnel current is interpreted as a bias dependence of the tunnelling gap while, at smaller distance, the bias dependence of the surface recombination velocity plays a dominant role.

The present model can be used as a basis for the interpretation of future spin-dependent tunnel photocurrent data.

**Acknowledgements**

The authors acknowledge X. Wallart for the epitaxial growth. This work was partially funded by the ANR SPINJECT-06-BLAN-0253.

Appendix A: Expressions of $v_d$ and $N_0$.

The charge diffusion equation is of the form

$$D\frac{\partial^2 n}{\partial z^2} - \frac{n}{\tau} + g\alpha\exp(-\alpha z) = 0 \tag{A1}$$

where $g$ is the density of impinging photons per unit time, $\alpha$ is the light absorption coefficient, $\tau$ is the bulk photoelectron lifetime and $D$ is the diffusion constant. For a planar sample of thickness $\ell$, the general solution of Eq. (A1) is

$$n_+ + n_- = Ae^{-z/L} + Be^{z/L} + \frac{g\alpha\tau}{1-(\alpha L)^2}e^{-\alpha z} \tag{A2}$$

where $L = \sqrt{D\tau}$ is the electron diffusion length. Using $\left.\frac{\partial n}{\partial z}\right|_0 = S'n(0)$ and $\left.D\frac{\partial n}{\partial z}\right|_{\ell-W} = -Sn_0$ as boundary conditions, one finds that the electronic concentration $n_0$ at $z=\ell-W$ and the photocurrent are given by Eq. (5) and Eq. (6), respectively, where

$$N_0 = \frac{g\alpha\tau}{(\alpha L)^2 - 1}\frac{\mu\alpha L - \upsilon + (S'L/D)[\mu - \upsilon\alpha L]}{(S'L/D)Ch(\ell/L) + Sh(\ell/L)} \tag{A3}$$

$$v_d = \frac{D}{L}\frac{(S'L/D)Ch(d/L) + Sh(d/L)}{Ch(d/L) + (S'L/D)Sh(d/L)} \tag{A4}$$

where the quantities $\mu$ and $\upsilon$ are given by

$$\mu = 1 - e^{-\alpha\ell}Ch(\ell/L) \qquad \upsilon = e^{-\alpha\ell}Sh(\ell/L) \tag{A5}$$

For an unpassivated rear surface, one has $S'L/D \gg Th(\ell/L)$ and $S'L/D \gg [Th(\ell/L)]^{-1}$ and

$$v_d = (D/L)[Th(\ell/L)]^{-1} \tag{A6}$$

Further assuming that $\alpha L \gg 1$ and neglecting for a large value of $\alpha\ell$ the light absorption at the front surface, one has $\mu \approx 1$ and $\upsilon = 0$ and

$$N_0 \approx \frac{g\tau}{LSh(\ell/L)} \tag{A7}$$

Appendix B: Tunnel current from the conduction band

We first write, for a given energy $\varepsilon_c$ above the bottom of the conduction band, the conservation of the perpendicular electronic momenta $k_\perp$, in the conduction band $i\kappa_\perp$ in the tunnel gap and $k'_\perp$ in the metal. $\kappa$ is related to the electron mass $m$ by $\hbar^2\kappa^2/2m = \overline{\Phi} - \varepsilon_{c\perp}$ where $\varepsilon_{c\perp} = \hbar^2 k_\perp^2/2m$ is the fraction of the energy $\varepsilon_c$ above the bottom of the conduction band corresponding to a kinetic energy perpendicular to the surface. The spatially-averaged value of the tunnel barrier $\overline{\Phi}$ for electrons at the bottom of the conduction band depends on bias.[20] Neglecting image charge effects, it is given by

$$\overline{\Phi} = [\Phi_m + \chi - E_b + qV]/2 \qquad (B1)$$

where $\Phi_m$ and $\chi$ are respectively the metal work function and the semiconductor affinity and $E_b$ is the energy of the bottom of the conduction band at the surface. The momentum $k'_\perp$ is obtained by expressing conservation of energy and of parallel momentum. Assuming that $\exp(-2\kappa d) \ll 1$, one finds that the tunnel probability is proportional to $G(\varepsilon_c)\exp(-2\kappa d)$ where

$$G(\varepsilon_c) = \frac{k}{k'} \frac{k_\perp^2}{(k_\perp + k'_\perp)^2 k_\perp^2 + (\kappa + k_\perp k'_\perp/\kappa)^2} \qquad (B2)$$

The tunnel current also depends on the product $K^*\rho_m(E)n_s(\varepsilon_c)W(\varepsilon_c)$ where $K^*$ is a constant, $\rho_m(E)$ is the metallic density of states at the corresponding energy $E$. Here $n_s(\varepsilon_c)$ and $W(\varepsilon_c)$ are the volume density of the electron concentration and the width of the depletion zone at energy $\varepsilon_c$. Thus, $n_s(\varepsilon_c)W(\varepsilon_c)$ is a concentration per unit area. One has finally, to first order in $\varepsilon_c/\overline{\Phi}$,

$$J_{tb} = K^* n_s \exp(-2d\sqrt{\overline{\Phi}}/d_0) \int_0^{\varphi_b} \rho_m(E) W(\varepsilon_c) G(\varepsilon_c) \rho(\varepsilon_c) \exp(\frac{\varepsilon_{c\perp} d}{d_0\sqrt{\overline{\Phi}}} - \frac{\varepsilon_c}{kT}) d\varepsilon_c \qquad (B3)$$

where $d_0 = \hbar/\sqrt{2m}$. For simplicity we only retain here the electrons which have the largest contribution to the integral of Eq. (B3). Because $W(\varepsilon_c)G(\varepsilon_c)$ increases with $\varepsilon_c$ and because of quantization of electronic states near the surface, the energy of these electrons is non-zero and will be written $f\varphi_b$ where $f$ is quite generally larger than $f^*$ defined in Eq. (9). Using Eq. (1) and Eq. (8), one finds $\exp(-\frac{\varepsilon_c}{kT}) \approx \exp\left[-\frac{f\varphi_b}{kT}\right] = \left[\frac{qSn_0}{A^{**}T^2}\right]^f$. The first exponential factor in the integral of Eq. (B3) is written $\frac{\alpha \varepsilon_c d}{d_0 \sqrt{\overline{\Phi}}}$ where

$$\alpha = \varepsilon_{c\perp}/\varepsilon_c \tag{B4}$$

Since the barrier value will be found weakly dependent on both bias and light excitation power, the product $W(\varepsilon_c)G(\varepsilon_c)\rho(\varepsilon_c)$ will be taken as constant and incorporated into the multiplicative constant, thus writing

$$K_b = K^* W(\varepsilon_c)G(\varepsilon_c)\rho(\varepsilon_c) \tag{B5}$$

Eq. (14) is finally obtained by developing $\sqrt{\overline{\Phi}}$ to first order in $qV$.

# Figures

Fig. 1: Metal-semiconductor structure excited by above bandgap light from the rear and for a positive bias *V* applied to the metal. Also shown are the surface density of states, peaking at midgap, and the energy difference $\Delta\varphi$ between the electron quasi Fermi level $E_{Fn}$ at the surface and the Fermi level $E_{F0}$ far from the junction in equilibrium. The shaded areas are the surface states lying between the electron and hole quasi Fermi levels (for which the energy difference is $qV_s$). Also shown are the metal work function, the semiconductor affinity, and the photocurrent ($J_p$) and Schottky current ($J_s$).

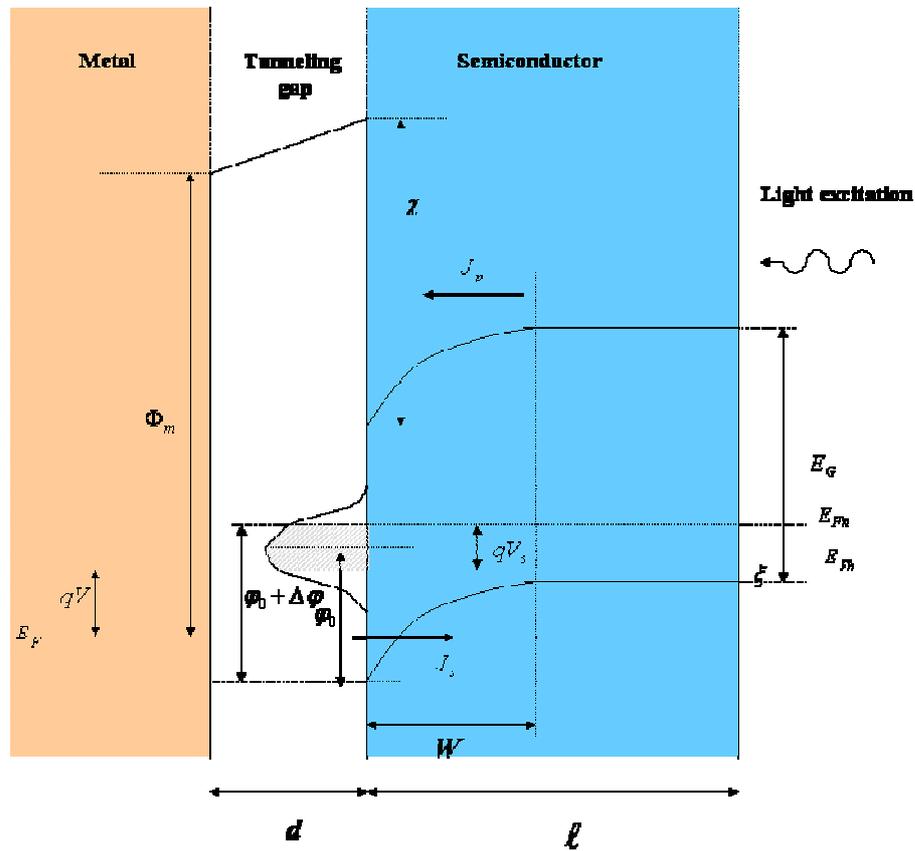

Fig. 2: Measured dark current versus bias for (a) $I_{set}=$ -2500, (b) -2000, (c) -1000, (d) 0, (e) 500, (f) 1500, (g) 2000 and (h) 3000. The exponential bias dependence of the current for a forward (positive) bias gives the ideality factor, which decreases up to $I_{set}=0$ and stays

approximately constant for larger values of $I_{set}$. The curves were rigidly shifted for clarity by a factor (d) 2, (e) 4, (f) 8, (g) 16 and (h) 30.

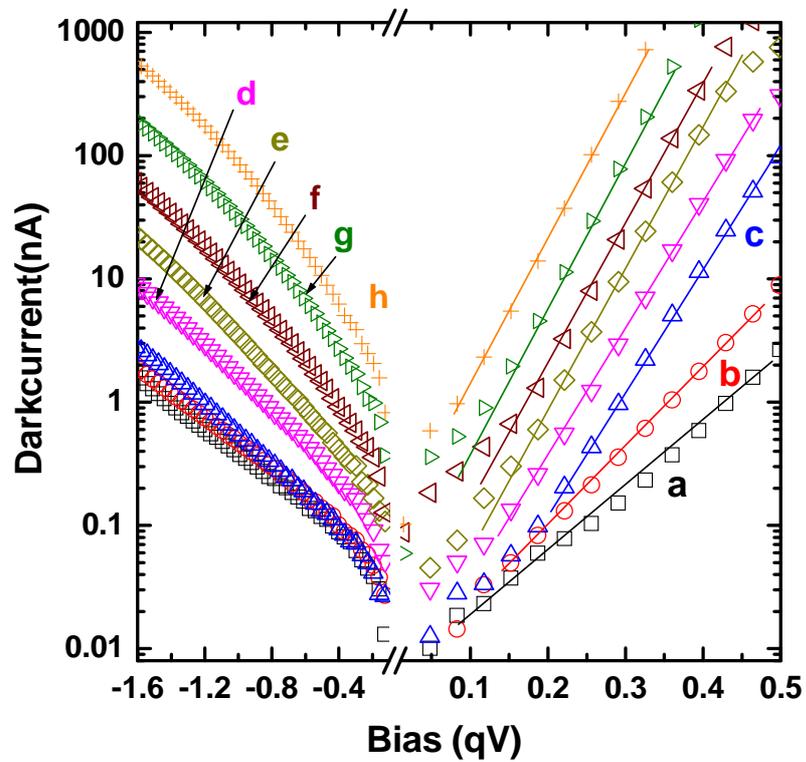

Fig. 3 : Tunnel photocurrent versus bias, defined as the additional tunnel current under light excitation. For (a) $I_{set}$= -2500, (b) -2000, and (c) -1000, the dependence at reverse (negative) bias is exponential. Progressive departure from purely exponential behavior occurs for (d) $I_{set}$= 0, (e) 500, (f) 1500, (g) 2000 and (h) 3000.

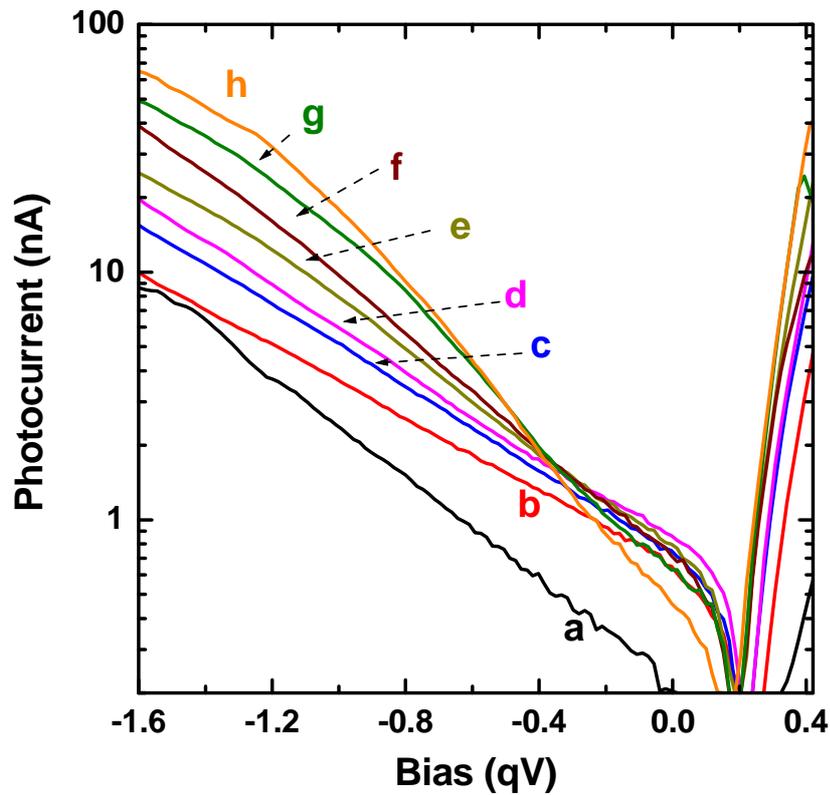

Fig.4. Atomic force between the cantilever and metal, as a function of $I_{set}$ measured in the same experiment as the tunnel currents. For negative values of $I_{set}$, the atomic force is approximately constant and taken as zero. As shown inset, mechanical contact occurs for positive values of $I_{set}$ situated between 0 and +1500. Here a clear mechanical bistability gives rise to two distinct values of the atomic force.

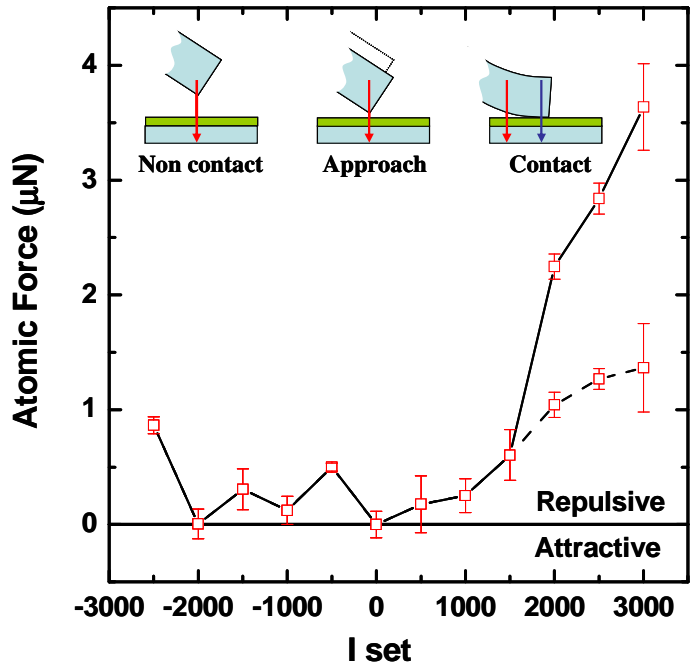

Fig. 5: Measured photocurrent versus light power at an applied bias of -1.5V for $I_{set}$= -3000,(a) -2000,(b) -1000,(c) and (3000).d Also shown for reference is a power law of exponent 0.5.

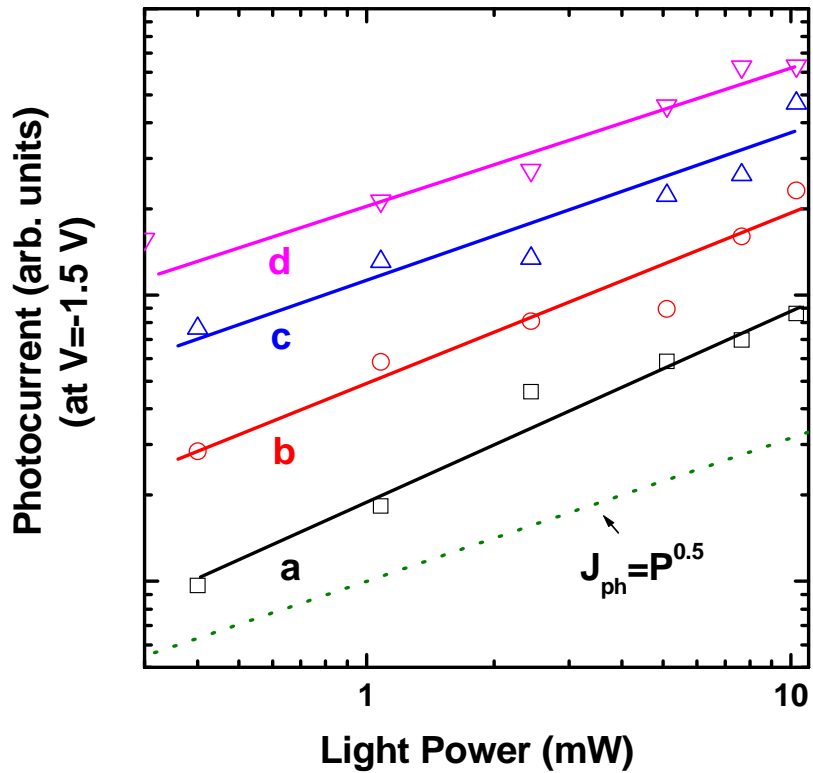

Fig. 6: Corrected bias dependences of the dark and tunnel photocurrents in mechanical contact. The curves correspond to (a) $I_{set} = 0$, (b) 1500, (c) 2000, (d) 2500 and (e) 3000 and were multiplied for clarity by a factor 2 for Curve c in the dark, 4 for Curve d and 10 for Curve e. The bias dependence of the curves depends very little on $I_{set}$ and, along with the improved exponential character at forward (positive) bias, shows that each curve corresponds to a constant tunnel distance.

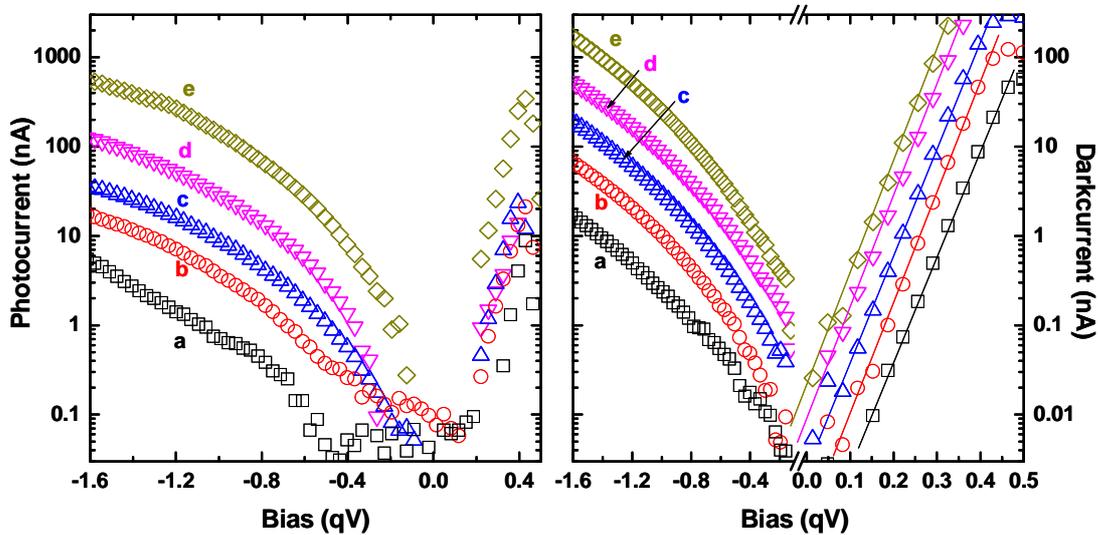

Fig. 7: Summary of the experimental bias dependences of the tunnel photocurrent (left panel) and dark current (right panel). The out-of-contact dependences have been obtained for (a) $I_{set}$=-2500, (b) -2000, and (c) -1000. The contribution of the contact to the bias dependence (d) was taken from the data obtained for $I_{set} = 3000$ after correction. For clarity the data for the tunnel photocurrent were multiplied by (b) 2, (c) 4 and (d) 10, while the multiplication factors for the dark current were (b) 2, (c) 3, and (d) 4. Note that the dark current is about one order of magnitude smaller than the tunnel photocurrent. Lines show the calculated currents found using the parameters in Table 2.

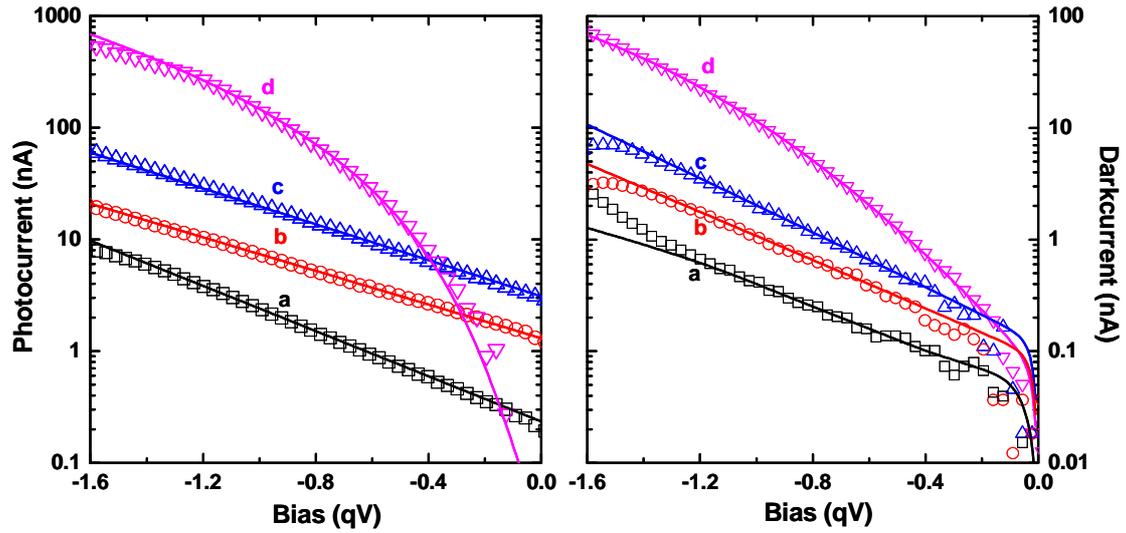

Fig. 8: Tunnel photocurrents from surface states and from the valence band calculated for $I_{set}$ = -2500 using Eq. (19) and Eq. (20) respectively. The other fixed parameter values that have been used are shown in Table 2. None of these currents is able to account for the data.

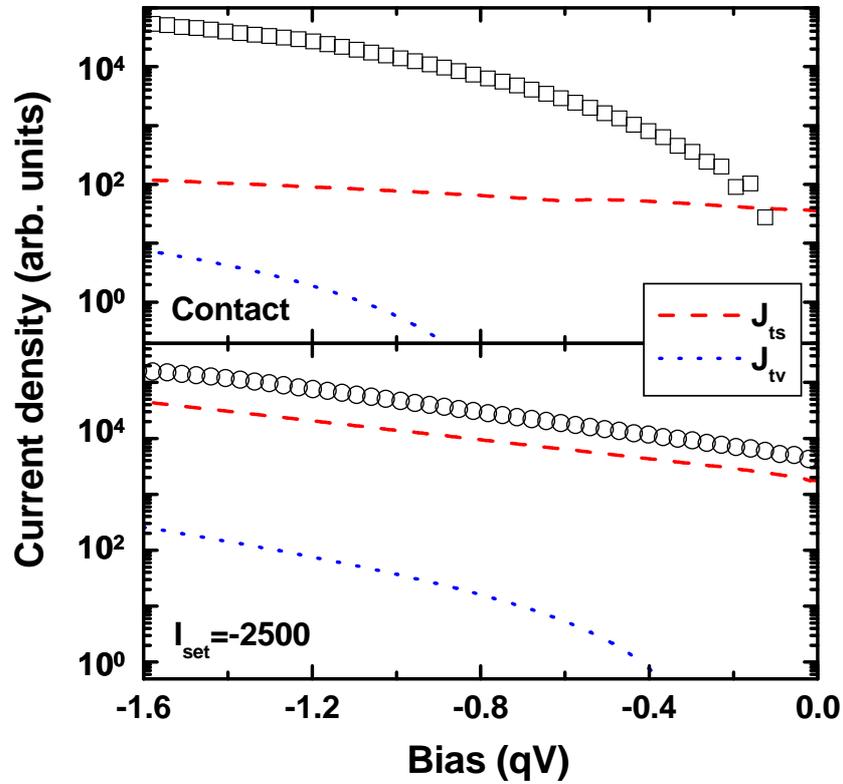

Fig. 9: Explanation for the distinct bias dependences before and after mechanical contact. This figure shows the measured bias dependence of the photocurrent for $I_{set}$=-2500 and for mechanical contact as well as the calculated bias dependence of the second and third terms of Eq. (14) as calculated using the parameter values shown in Table 2. Before contact the bias dependence of the tunnel photocurrent is determined by that of the tunnel barrier. After contact the bias dependence of the surface recombination velocity plays a dominant role.

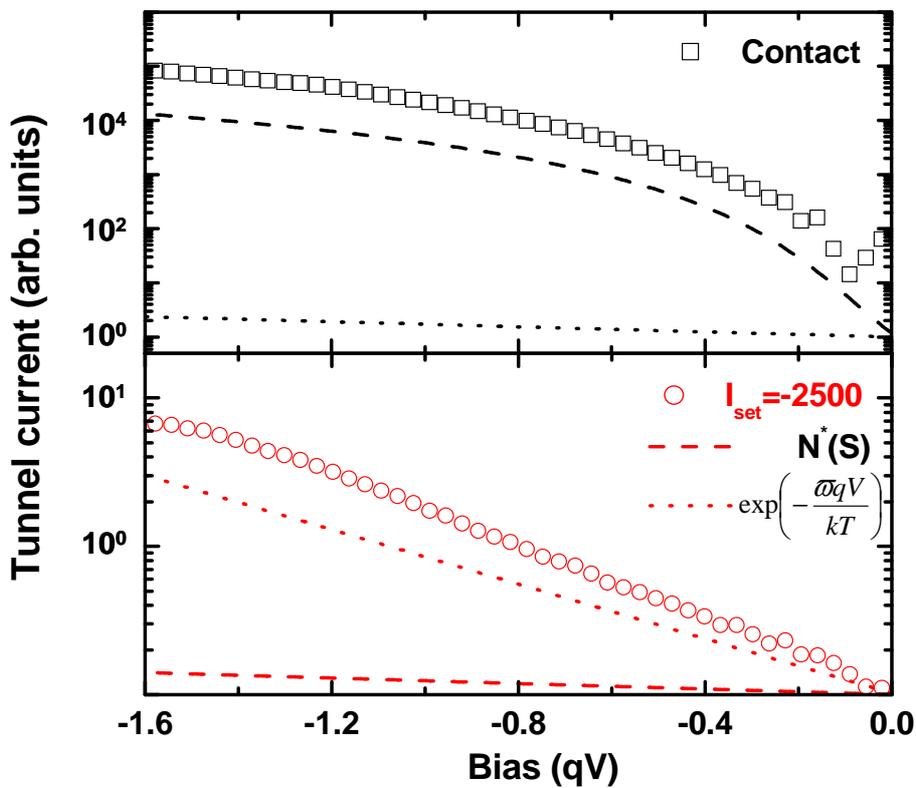

Fig. 10: Calculated dependence on distance of the dielectric constant out of contact, assuming that the metal is covered by a thin layer of thickness smaller than the tunnel distance. The data points correspond to the values used in the analysis of the out of contact curves. The inset shows various parameters used in the calculation as a function of $\varepsilon_0/C_m$. Image charge effects and tunnelling of majority carriers contribute to an ideality factor $(1-\alpha_d)^{-1}$ and only play a significant role in contact.

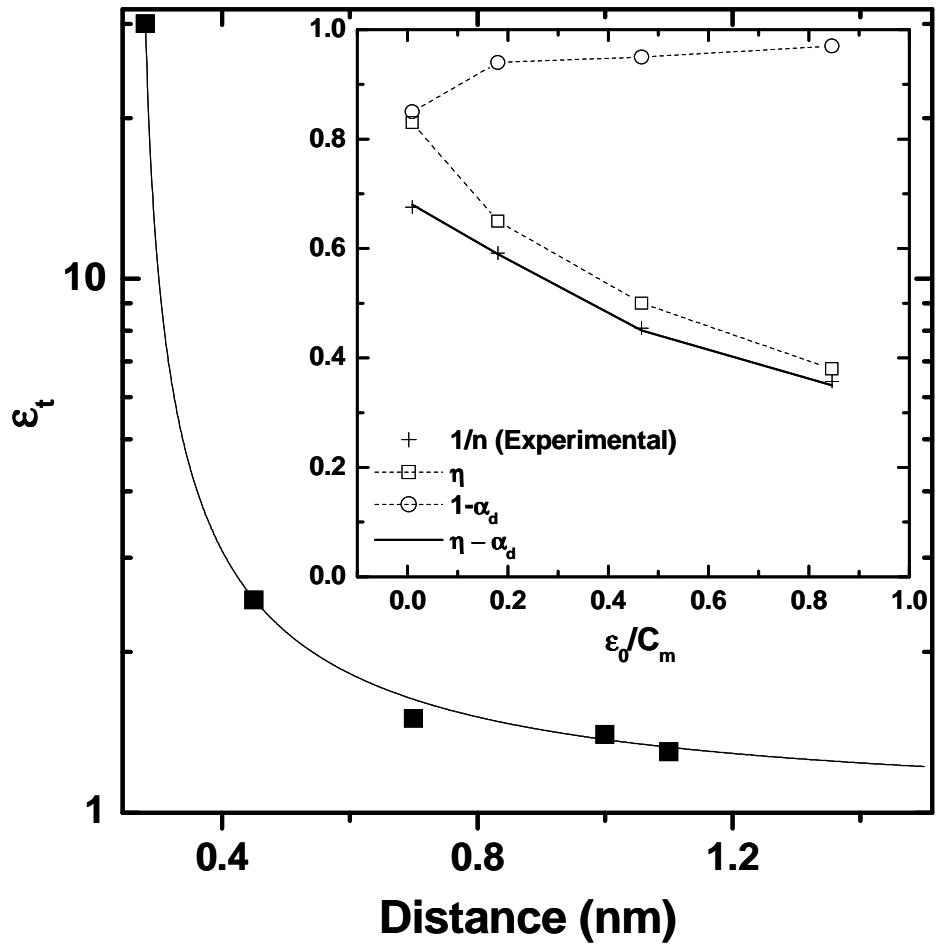

Table 1: Experimentally measured tunnel currents in contact were corrected by subtracting a fraction α of the tunnel current corresponding to the largest value of $I_{set}$ out of contact. This table shows the values of α as a function of $I_{set}$. Bistability of the atomic force (see Fig. 4) in contact is correlated with two distinct values of $\alpha$ shown here for $I_{set} = 3000$.

| $I_{set}$ | 3000 | 2500 | 2000 | 1500 | 0 |
|---|---|---|---|---|---|
| α | 1<br>0.7 | 0.8 | 0.85 | 1 | 1 |

Table 2: Values of parameters used for the analysis of the Curves of Fig. 7.

*Common parameters*

| Parameter | Value |
|---|---|
| $S_0 / v_d$ | 62.5 |
| $J_{sat}$ (A/m$^2$) | 6x10$^{10}$ |
| $N_A$ (m$^{-3}$) | 10$^{24}$ |
| $N_T(0)$ (eV$^{-1}$.m$^{-2}$) | 6x10$^{18}$ |
| $N_T^D(0)$ (eV$^{-1}$.m$^{-2}$) | 6x10$^{18}$ |
| $\sigma$ (eV) | 0.20 |
| $N_0$ (m$^{-3}$) | 2x10$^{22}$ |
| $f$ | 0.38 |

*Adjustable parameters*

| $I_{set}$ | Contact | -1000 | -2000 | -2500 |
|---|---|---|---|---|
| $\varepsilon_0/C_m$ (nm) | 0.009 | 0.18 | 0.47 | 0.85 |
| ω | 0.011 | 0.017 | 0.027 | 0.043 |
| $\alpha_d'$ ($10^{-3} V^{-1}$) | 30 | <2 | <2 | Irrelevant |

# References

+ corresponding author. e-mail address: daniel.paget@polytechnique.fr

[1] I. Zutic, J. Fabian and S. Das Sarma, Rev. Mod. Phys. **76** 323-410 (2004).

[2] D.T. Pierce Physica Scripta **38**, 291 (1988)

[3] M. Bode Rep. Prog. Phys. **66**, (2003), 523.

[4] M. W. J. Prins, R. Jansen, R. H. M. Groeneveld, A. P. van Gelder, and H. van Kempen, Phys. Rev. B **53**, 8090 (1996)

[5] Y. Suzuki, W. Nabhan, and K. Tanaka, Appl. Phys. Lett. **71**, 3153, (1997).

[6] R. Jansen, R. Schad and H. Van Kempen, J. Mag and Mag. Mater. **198**, 668, (1999)

[7] D. Paget, J. Peretti, A. C. H. Rowe, G. Lampel, B. Gérard, S. Bansropun, French Patent N° 05 05394,(2005).

[8] D. Vu, R. Ramdani, S. Bansropun, B. Gérard, E. Gil, Y. André, A. C. H. Rowe, D. Paget, J. Appl. Phys, **107**, 093712, (2010).

[9] E. H. Rhoderick *Metal-semiconductor contacts* Clarendon (Oxford) 1978.

[10] L. Kronik, and Y. Shapira, Surf. Sci. Rep. **37**, 1, (1999).

[11] S. Grafström, J. Appl. Phys. **91**, 1717, (2002).

[12] A. C. H. Rowe and D. Paget, Phys. Rev. B **75**, 115311 (2007).

[13] R. Jansen, M. W. J. Prins, and H. van Kempen, Phys. Rev. B **57**, 4033 (1998).

[14] M. W. J. Prins, R. Jansen, and H. van Kempen, Phys. Rev. B **53**, 8105 (1996).

[15] J. Reichman, Appl. Phys. Lett. 36, 574, (1980).

[16] S. Arscott, E. Peytavit, D. Vu, A.C.H. Rowe and D. Paget, J. Micromech. Microeng. **20**, 025023 (2010)

[17] W. G. Gärtner, Phys. Rev **116**, 84, (1959)

[18] D. Aspnes, Surf. Sci, **132**, 406, (1983).

[19] E. W. Kreutz, Phys. Stat. Sol. (a) **56**, 687, (1979).

[20] J. G. Simmons, J. Appl. Phys. **34**, 1793, (1963).

[21] D. A. Papaconstantopoulos, *Handbook of the Band Structure of Elemental Solids,* Plenum Press, New York, 1986.

[22] J. He, M. Chan and Y. Wang, IEEE trans. El. Dev. **53**, 2082, (2006).

[23] C. G. B. Garret and W. H. Brattain, Phys. Rev **99**, 376, (1955).

[24] R. A. Smith, *Semiconductors,* Second Edition, Cambridge University Press, Cambridge, 1978.

[25] C. H. Henry, R. A. Logan, and F. R. Merritt, J. Appl. Phys. **49**, 3530, (1978).

[26] L. Landau and E. Lifchitz, *Quantum mechanics*, Mir, Moscow, 1966.

[27] P. Prod'homme, F. Maroun, R. Cortes, and P. Allongue, Appl. Phys. Lett. **93**, 171901, (2008).

[28] Ş. Katartaş, and Ş. Altındal, Mat. Sci and Eng. B **122**, 133, (2005).

[29] $J_{ts}$ and $J_{tv}$ are small with respect to $J_{tb}$ because their respective tunnel barriers are larger than for conduction electrons, or possibly because of their relatively small tunnel matrix elements and coherence length $\ell_c$ for $J_{tv}$ in Eq. (20). In agreement with these conclusions, the tunnel current from the conduction band of n-type GaAs is known to be larger than that from the valence band [R. M. Feenstra, Phys. Rev. B **50**, 4561, (1994)]. Moreover, no tunnel current from defects is found on oxygen covered GaAs .[R. M. Feenstra, and J. A. Stroscio, J. Vac. Sci. Technol. B 5, 923 (1987)].

[30] N. L. Dmitruk, O. Yu Borkovskaya, and O. V. Fursenko, Vacuum, **50**, 439, (1998).

[31] N. C. Quach, N. Simon, I. Gérard, P. Tran Van and A. Etcheberry, J. Electrochem. Soc. **151**, C318, (2004)

[32] H. C. Card and E. H. Rhoderick, J. Phys. D **4**, 1589, (1971).

[33] This approach neglects the tunnel current from the valence band $J_{tv}$. At short distance, the semiconductor band structure follows the motion of the metal Fermi level ($V_s = V$) so that

$J_{tv}$ =0 since valence band states lie below the metal Fermi level. At large distance, one would expect in the same way as in Fig. 8, appearance of $J_{tv}$ at a bias of the order of -0.6V. Such effect is not observed experimentally, probably because of the smallness of the tunnel matrix element or of the coherence length in the valence band which as seen in Eq. (18) appears in the two-dimensional density of states in the valence band

[34] The presence of two distinct types of surface states and the bias dependence of the barrier $\varphi_0^*$ have not been taken account under light excitation. This is reasonable since i) photoelectron capture processes increase the kinetics of establishment of equilibrium with the semiconductor, ii) because of the photovoltage, the correction term, proportional to $\Delta\varphi - qV_s$ is smaller under light excitation than in the dark.